\documentclass[twocolumn,showpacs,preprintnumbers,amsmath,amssymb,prl,floatfix]{revtex4}

\usepackage{graphicx}

\usepackage{color}
\usepackage{dcolumn}
\usepackage{bm}
\usepackage{longtable}
\usepackage{epsfig}
\usepackage{times}

\begin{document}

\title{Single-Photon Entanglement in the keV Regime via Coherent Control of Nuclear Forward Scattering}

\author{Adriana \surname{P\'alffy}}
\email{Palffy@mpi-hd.mpg.de}

\author{Christoph~H. \surname{Keitel}}

\author{J\"org \surname{Evers}}
\email{Joerg.Evers@mpi-hd.mpg.de}

\affiliation{Max-Planck-Institut f\"ur Kernphysik, Saupfercheckweg~1, 
69117 Heidelberg, Germany}

\date{\today}

\begin{abstract}
Generation of single-photon entanglement  is discussed in nuclear forward scattering. Using successive switchings of the direction of the nuclear hyperfine magnetic field, the coherent scattering of photons on nuclei is controlled such that two signal pulses  are generated out of one initial pump pulse. The two time-resolved correlated signal pulses have different polarizations and energy in the keV regime.
Spatial separation of the entangled field modes and extraction of the signal  from the background can be achieved with the help of state-of-the-art x-ray polarizers and  piezoelectric fast steering mirrors.

\end{abstract}






\maketitle


Almost 75 years after its unexpected birth in connection with the  Einstein-Podolsky-Rosen paradox~\cite{EPR}, the concept of entanglement remains one of the most fascinating in physics. Quantum  entanglement plays a key role in many of the most interesting aspects and applications of quantum information and  computing. While fundamental questions related to local realism, hidden variables, Bell's inequalities or the completeness of quantum mechanics as a theory gravitate around  entanglement~\cite{Nielsen}, achievement of teleportation of a quantum state has revolutionized its possible applications~\cite{TelePRL,TeleReview}. Surprisingly, teleportation could be realized not only by using entangled pairs of photons or particles, but also with field modes entangled by one single photon~\cite{Lombardi_Babichev}. In such a single-photon entangled state
\begin{equation}
\label{spe}
|\Psi\rangle = |1\rangle_A|0\rangle_B + |0\rangle_A|1\rangle_B\,,
\end{equation}
the spatially or temporally separated field modes $A,B$ are entangled. The peculiarity that only one photon is involved in the entanglement process has generated some debate~\cite{sph_entangl}. 

 Whether spins of trapped ions or nuclei, single quantum oscillators, photons in an optical cavity, or the variety of different single-photon sources,  diversity characterizes the applications of entanglement and there is no  preferred test system. 
The parameter regime of entanglement applications has been until now however limited. The experimental realization of quantum  teleportation, for instance, even when following different  ingenuous routes and setups, always involved optical photons as teleported qubits. Probing entanglement in new parameter regimes, for instance with keV photons from nuclear transitions, has not yet been realized. At present we are only aware of  the proposal in Ref.~\cite{Unruh} of a tabletop source for pairs of entangled keV photons  as signatures of the Unruh effect, created by electrons moving in a strong periodic electromagnetic field.
When properly controlled, at best coherently, nuclei could provide a source of entanglement in the x-ray regime. 
Due to their inner structure, nuclei not only offer the possibility to explore a new energy regime, but also to extend to a new degree of complexity.

Coherent control of nuclear excitations has been a long-time goal in nuclear physics, 
as it is related to a number of promising applications such as nuclear 
quantum optics~\cite{nqo,isomers,nuclear_laser,Odeurs}, including isomer triggering~\cite{isomers} or nuclear lasers~\cite{nuclear_laser,Odeurs}. Nuclear forward scattering (NFS) of synchrotron radiation (SR)~\cite{NFSReviews}  is one promising experimental setup in which coherent control has been successfully applied. Ref.~\cite{Shvydko_MS} reported controlling the coherent decay of nuclear excited states in NFS by switching the direction of the hyperfine magnetic field in the nuclear target. The coherence occurs due to the formation of an excitonic state whose decay can be controlled via interference effects, similarly to the  electromagnetic induced transparency in quantum optics~\cite{RMP,Odeurs}. Thus, the magnetic switching experiment in Ref.~\cite{Shvydko_MS} is a promising experimental conceptual proof  of coherent control in nuclei. 

In this Letter, we present a coherent control scheme to generate keV single-photon entanglement in a NFS setup. The coherent scattering of the SR pump pulse can be  controlled by a sequence of magnetic field switchings such that two time-resolved entangled coherent decay pulses with different photon polarizations are emitted. Since the SR pulse typically creates only one (and more often no) nuclear resonant excitation in the target~\cite{PotzelPRA63}, with high probability only one photon will be emitted in either of the two entangled field modes, with no way of knowing beforehand in which one. Because of their different polarizations, the two entangled modes can be spatially separated with the help of state-of-the-art x-ray polarizers to give a state as in Eq.~(\ref{spe}). A method to extract the signal photons from the strong background in the NFS setup with  piezoelectric fast steering x-ray mirrors is discussed.


Resonant scattering  of SR on a nuclear ensemble,  such as identical nuclei in a crystal lattice, occurs via an intermediate excitonic state~\cite{Tramell_book_AK}. This collective nuclear excited state decays coherently in the forward direction, giving rise to NFS, and in the case of nuclei in a crystal also at the Bragg angle~\cite{NFSReviews}. A further consequence of the  excitonic state is  the speed up of the coherent decay~\cite{Kagan_JPC_vanB_PRL59,Hastings}. The coherent decay channel thus becomes considerably faster than the spontaneous (or incoherent) one, which is characterized by the natural lifetime of a single nucleus. This opens the  possibility to manipulate the resonant scattering via the coherent nuclear decay.

\begin{figure}
\begin{center}
\includegraphics[width=0.4\textwidth]{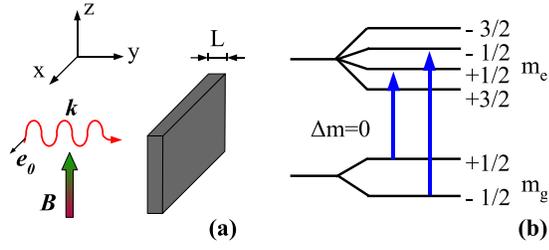}
\caption{\label{geometry} (Color online) Scattering geometry (a), and hyperfine level scheme of the  $^{57}\mathrm{Fe}$ transition (b).
 The incident radiation has the wavevector $\vec{k}$ parallel to the $y$ axis and the polarization $\vec{e}_0$ parallel to the $x$ axis. The initial hyperfine magnetic field $\vec{B}$ is oriented parallel to the $z$ axis. In this setup the SR pulse can only induce  $\Delta m=0$ nuclear transitions. }
\end{center}
\end{figure}

We consider the traditional $M1$ M\"ossbauer excitation of $^{57}\mathrm{Fe}$ nuclei from the ground state to the first excited state at 14.413 keV. The natural lifetime of the excited state is $\tau_0=141.1$~ns. In the presence of a strong magnetic field, the $^{57}\mathrm{Fe}$ nuclear ground and excited states (of spins $I_g=1/2$ and $I_e=3/2$, respectively) split into hyperfine levels, as shown schematically in  Fig.~\ref{geometry}. Depending on the geometry of the setup and the polarization of the incident SR, some of the six possible hyperfine transitions will be excited by the SR  pump pulse, with $\Delta m=m_e-m_g\in \{0, \pm 1\}$, where $m_g$ and $m_e$ are the projections of the nuclear spin on the quantization axis. Coherent scattering implies that  initial and  final states coincide. Fulfilling this requirement is also crucial for the final-state indistinguishability necessary for entanglement.

\begin{figure}
\begin{center}
\includegraphics[width=0.5\textwidth]{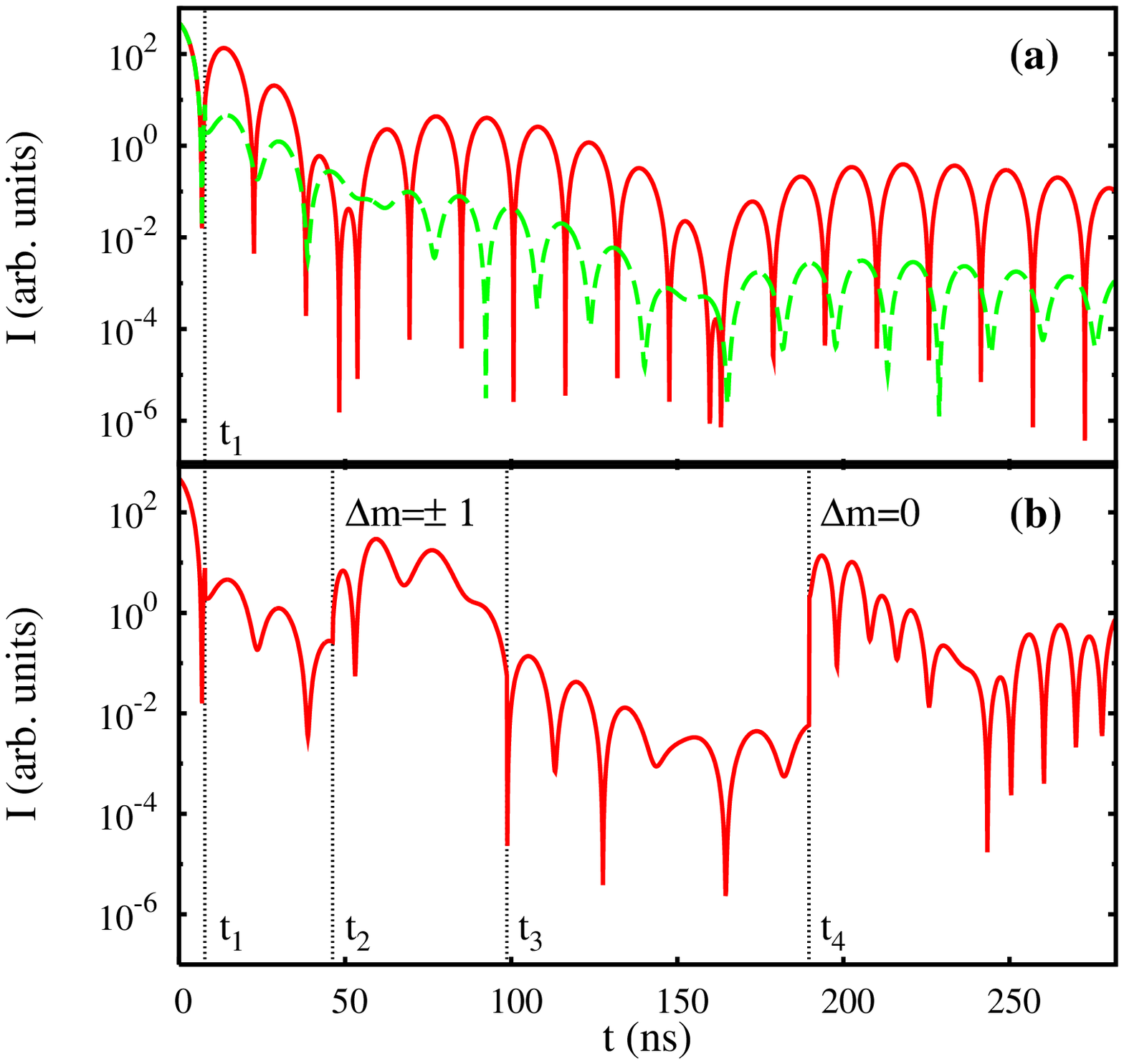}
\includegraphics[width=0.5\textwidth]{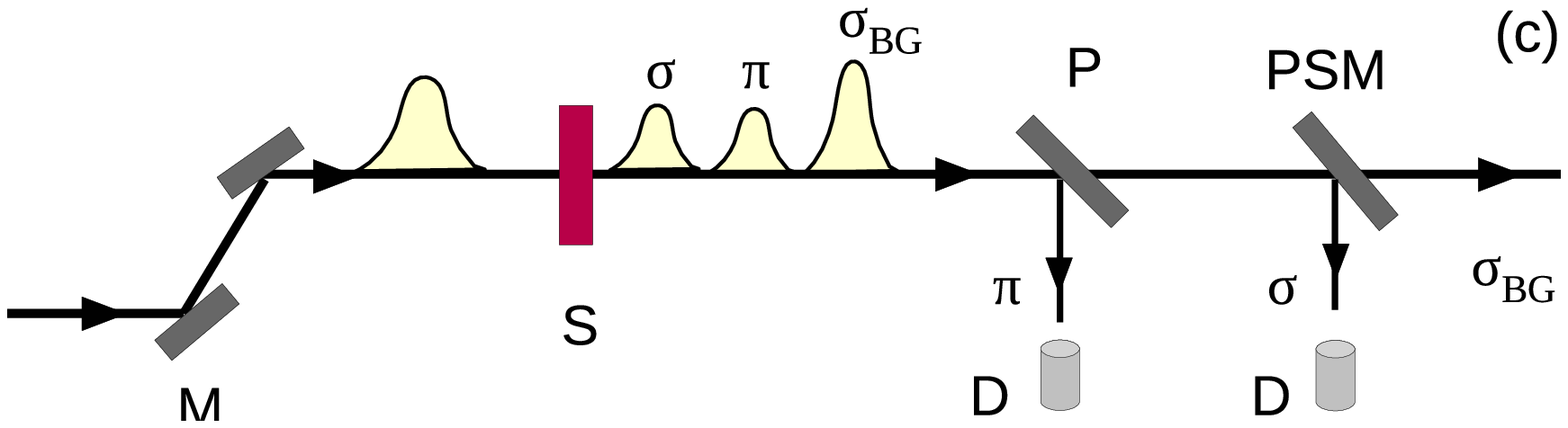}
\caption{\label{intensity} (Color online) Calculated time spectra for the coherent nuclear decay: (a) unperturbed (full line) and suppressed  by a magnetic switching at time $t_1$ (dashed line); (b)  with four magnetic switchings, which create two light pulses of different polarization. (c) Sketch of the discussed setup. The SR pulse is monochromatized (M) before it reaches the sample (S). The $\pi$-polarized pulse is selected from the forward response by a polarizer (P), and the $\sigma$-polarized pulse is extracted from the background ($\sigma_{BG}$) with the help of a piezoelectric fast steering mirror (PSM). D are detectors. Pulses not drawn to scale.} 
\end{center}
\end{figure}

The scattering geometry is shown in Fig.~\ref{geometry}(a). The incident SR is monochromatized up to meV energies~\cite{Hastings} in order to eliminate part of the strong background. As in~\cite{Shvydko_MS}, we chose the $z$ axis along the hyperfine magnetic field $\vec{B}$ and the $y$ axis parallel to the incident photons. The nuclei are embedded in a crystal plate of thickness $L$ in the $y$ direction, with $y=0$ at the crystal entrance surface. The incident beam has electric field polarization $\vec{e}_0$ parallel to the $x$ axis as shown in Fig.~\ref{geometry}(a). Thus, magnetic $\sigma$-transitions with $\Delta m=m_g-m_e=0$ are excited.  The amplitude $\vec{E}(y,t)$ of the radiation pulse caused by the coherent forward scattering can be calculated using standard methods from the wave equation~\cite{Shvydko_theory,Motif}
\begin{equation}
\frac{\partial\vec{E}(y,t)}{\partial y}=-\sum_{\ell} K_{\ell} \vec{J}_{\ell}(t)\int_{-\infty}^t d\tau \vec{J}^{\, \dagger}_{\ell}(\tau)\cdot \vec{E}(y,\tau)\, .
\label{wave_eq}
\end{equation}
We use the nuclear transition currents $\vec{j}_{\ell}(\vec{k})$ 
and their matrix elements $\vec{J}_{\ell}(t)$ and $\vec{J}^{\, \dagger}_{\ell}(t)$ 
corresponding to the nuclear decay and excitation steps, respectively, as defined in Ref.~\cite{Shvydko_MS}.
 The summation index $\ell$ runs over the magnetic sublevels $m_e$ and $m_g$, and $K_{\ell}$ are the coefficients defined in \cite{Shvydko_MS}.
An example of the total intensity $I(t)$ is shown by the full red line in Fig.~\ref{intensity}(a). 
Two beat patterns can be observed: the quantum beat due to the interference between the two transitions driven by the SR pulse, given by the sum over $\ell$ in Eq.~(\ref{wave_eq}), and the dynamical beat due to multiple scattering contributions.

Magnetic switching is achieved by changing abruptly the direction of the hyperfine magnetic field at the nuclei after the excitation. This is possible in crystals that allow for fast rotations of the strong crystal magnetization via weak external magnetic fields. The switching experiment in Ref.~\cite{Shvydko_MS} was facilitated by $^{57}\mathrm{FeBO}_3$, a canted antiferromagnet with a plane of easy magnetization parallel to the (111) surfaces. Initially, a constant weak magnetic field induces a magnetization parallel to the crystal plane surface and aligns the magnetic hyperfine field $\vec{B}$ at the nuclei. The magnetic switching is then achieved by a stronger, pulsed magnetic field in a perpendicular crystal plane, that rotates the magnetization by an angle $\beta$ and realigns the hyperfine magnetic field.  Because of the perfection of the crystal, the desired rotation of the magnetization occurs abruptly, over less than 5~ns~\cite{Shvydko_EPL}. 
A rotation of the hyperfine magnetic field leads to a new quantization axis $z'$, and therefore to new eigenvectors of the hyperfine Hamiltonian. This redistributes the original nuclear state populations into the new eigenstates.
Each transition $\ell$ between the original hyperfine levels is then transformed into a multiplet composed of all allowed transitions $\{\ell'\}$ projected onto the new quantum axis $z'$. 

The new currents $\vec{j}_{\ell'}(\vec{k})$ interfere, and depending on the switching time $t$ and the switching angle $\beta$, their interference can be constructive or destructive. Suppression or restoration, i.e., control of the coherent nuclear decay, can thus be achieved by optimizing the time and angle of switching. For instance, complete suppression of all possible transitions in the dominant first order scattering can be achieved by rotating the hyperfine magnetic field into a direction parallel to the incident radiation $\vec{k}$ at time $t_1=(n-1/2)\pi/\Omega_0$, where $n\in \mathbb{N}$ and $\Omega_0$ is the hyperfine energy correction for the $\Delta m=0$ transitions~\cite{Shvydko_MS}. An example is shown in Fig.~\ref{intensity}(a), calculated with a switching at $t_1=\pi/(2\Omega_0)$. Since coherent decay is almost completely suppressed, the excitation energy is stored inside the crystal. This storage is imperfect due to small non-zero contributions of the higher-order scattering terms.

Let us  now turn  to the switching scheme we employ to produce keV  single-photon entanglement. Initially, the hyperfine magnetic field $\vec{B}$ is oriented parallel to the $z$ axis, see Fig.~\ref{geometry}(a).  The SR pump probe reaches the sample at time $t=0$, producing the nuclear excitonic state.  At time $t_1=\pi/(2\Omega_0)$, the hyperfine magnetic field $\vec{B}$, originally oriented parallel to the $z$ axis, is switched in the direction of the incident radiation, so that complete suppression of the first-order scattering occurs due to destructive interference. The coherent decay in the time interval $0<t<t_1$ is not useful for our entanglement scheme and will be treated as background. 
The stored excitation can be released by rotating back the hyperfine magnetic field parallel to the $z$ axis. The second switching time $t_2$ is again crucial, as it allows for control of the polarization of the subsequently released photons.  By choosing a suitable $t_2$, only currents corresponding to the $\pi$-polarized component will contribute to the NFS signal, while the ones corresponding to $\sigma$-polarization will continue to interfere destructively. Our calculation shows that switching back to the initial direction of the magnetic field at $t_2=46$~ns, as shown in Fig.~\ref{intensity}(b), sets free only the $\Delta m=\pm 1$ frequency components, whereas the ones with $\Delta m=0$ are still suppressed. 
In order to separate the outcoming scattered light in two entangled keV photon pulses, two more  switchings are required. The coherent decay can be again almost completely suppressed by a rotation of the magnetic field to the direction of the incident radiation at $t_3=99$~ns. The switching time is chosen such that after $t_3$, all new currents interfere destructively. Finally, a rotation of the magnetic field back to its original direction along the $z$ axis at $t_4=190$~ns will release all remaining stored energy into photons with $\sigma$-polarization via the $\Delta m=0$ transitions. In total, after the initial response until $t_1$, the described switching sequence leads to an emission of the stored excitonic energy into two photon pulses of different polarization after $t_2$ and $t_4$, as shown in Fig.~\ref{intensity}(b).  

A similar separation of a single photon into two polarization modes could also be achieved otherwise, e.g., via scattering in the Faraday geometry~\cite{NFSReviews}. However, the advantage of our switching scheme is that the entangled photon pulses are emitted immediately after the switchings, in time windows which are substantially shorter than the nuclear decay time. This is an important requirement for possible applications, which are typically based on correlation of coincidence measurements.

The next step is to split the temporally separated pulses into distinct spatial modes; see Fig.~\ref{intensity}(c). 
Whereas the separation of the  $\pi$- and $\sigma$-polarized signal photon components into different spatial modes $A,B$ generates a single-photon entangled state as in Eq.~(\ref{spe})~\cite{vanEnk}, 
the spatial separation of signal photon and background is desirable to achieve a source of  keV  vacuum-entangled photons  as a resource for applications. 
The background arises close to $t=0$ in Fig.~\ref{intensity} from SR that traverses the sample unperturbed and due to prompt electronic scattering. The coherent nuclear decay before $t_1$ is also background in our entanglement scheme. In the following, we discuss in detail the spatial separation of the three pulses using x-ray optics techniques.

An obvious choice is to take advantage of the different polarizations of the two possible outcoming photons. Crystal reflections with Bragg angles near 45$^{\circ}$ are used in polarization-sensitive measurements  to overcome detector limitations in NFS~\cite{Siddons}. 
For the x-ray energy of the $^{57}\mathrm{Fe}$ first excited state, the Si(8~4~0) reflection with $\Theta_B=45.10^{\circ}$ is most suited for a polarized Bragg reflection~\cite{Toellner}. At this angle, the ratio of the integrated $\pi$ reflectivity to the integrated $\sigma$ reflectivity for a channel-cut crystal is  $10^{8}$~\cite{Toellner}.
In our case, the incoming SR radiation, the first background pulse emitted at $t\le t_1$ and the second signal pulse emitted at $t\ge t_4$  are  $\sigma$ polarized, while the first signal pulse is $\pi$ polarized. The $\pi$ pulse can be separated with a suitable polarizer, as shown in Fig.~\ref{intensity}(c).
 The total efficiency of the polarizer setup is limited by the 
very small angular acceptance $\Delta \Theta_B=10.6$ $\mu$rad of the Si(8~4~0) reflection \cite{Toellner}. Typical divergences at undulator beamlines of third generation facilities after monochromatization are in the range of 15 $\mu$rad vertical and 50 $\mu$rad horizontal, such that only a part (approximately 10~$\%$) of the scattered flux can be used~\cite{Roehl_hypint}.

The remaining task is to separate the $\sigma$-polarized entangled pulse component from the background radiation. Until now, several approaches to the background problem in NFS have been applied. A first successful ansatz relies on time gating to cut off the prompt background from the delayed signal, without spatial separation of signal and background. 
 Another approach relies on the so-called nuclear lighthouse effect that involves rotating the sample~\cite{NLE,Roehl_hypint}. In our case, however, due to the precise alignment of the magnetic fields that have to be applied to the crystal, a rotation of the sample is precluded.

Rather than reducing the background, it is favorable for quantum information applications to extract the $\sigma$-polarized entangled pulse component from the background radiation. 
Using Bragg reflections, silicon or sapphire x-ray mirrors can be designed for specific reflection energies and angles, with very small angular acceptances, on the order of $\mu$rad or less~\cite{Shvydko_book}. Within the Bragg angular acceptance range, the crystal will act as a x-ray mirror for radiation of specific energy, while it will transmit the radiation at different angles.
This suggests a setup in which an x-ray mirror for the 14.413 keV resonant energy of $^{57}\mathrm{Fe}$ is moved or rotated in and out of the reflecting position by a piezoelectric fast steering device. Alternatively, ultrafast, sub-ns piezo switches for x-ray SR which are not based on the motion, but rather on the lattice deformation of the mirror,  have already been used in other contexts~\cite{APL_JPD} and are promising devices in time-resolved measurements. In contrast to typical choppers, piezoelectric switching devices have the additional advantage of versatile synchronization with the incident SR.
Unlike in conventional NFS spectroscopy, the relevant  $\sigma$-polarized component of the entangled photon is emitted only after time $t_4$, while photons before $t_1$ should not be reflected. 
This leaves a maximum switching time from transmission to reflection of about 180~ns.
The combination of polarization-sensitive reflection and piezoelectric switching enables to extract the entangled photon pulse from the background, and to separate its different polarization components into separate spatial modes. Using this setup as a keV single-photon source, one can envisage a verification of the generated entanglement by performing  an experimental test of a single-particle version of Bell's inequality, as put forward in Ref.~\cite{Lee}.

Estimates can be carried out for the single-photon entanglement generation rate  considering the spatial separation scenario described above. The coherent decay at times $t<t_1$ is reducing the number of signal photons to $30\%$ of the initially excited nuclei. The losses due to incoherent decay are already included in the calculation of the coherent decay intensity. Furthermore, the incoherently emitted photons are not preferentially emitted in the forward directions so that we  neglected them as possible background for the single-photon entanglement scheme.
Using the polarizer to extract the $\pi$-polarized pulse, only about 10$\%$ of the incoming photons are kept due to the small angular acceptance of the polarizing crystal~\cite{Roehl_hypint}. Current experiments typically operate with incident photon fluxes of 10$^9$ photons/s  after the monochromator. Considering this incoming photon flux, which accounts for around 5$\times 10^5$ excited nuclei per second in the sample, we obtain a rate of approx. 15$\times 10^3$ vacuum-entangled photons per second using the polarizer to discern between the $\sigma$ and $\pi$ entangled field modes.
It should be noted, however, that since in our scheme the signal of interest is spatially separated from the background, the incoming SR flux can be increased without disturbing the detection as in time-gated setups. This primary idea may also have applications in other NFS setups.


\end{document}